\documentclass[12pt]{JHEP}
\usepackage{epsfig}

\def\lsim{\mathrel{\mathpalette\fun <}}
\def\gsim{\mathrel{\mathpalette\fun >}}
\def\fun#1#2{\lower3.6pt\vbox{\baselineskip0pt\lineskip.9pt
  \ialign{$\mathsurround=0pt#1\hfil##\hfil$\crcr#2\crcr\sim\crcr}}}
\newcommand{\beq}{\begin{equation}}
\newcommand{\eeq}{\end{equation}}

\title{The toes of the ultra high energy \\cosmic ray spectrum}

\author{Diego Harari$^a$, Silvia Mollerach$^b$ and Esteban Roulet$^b$
\\$^a$Departamento de F\'\i sica, FCEyN, Universidad de Buenos Aires
\\Ciudad Universitaria - Pab. 1, 1428, Buenos Aires, Argentina
\\$^b$Departamento de F\'\i sica, 
Universidad Nacional de La Plata\\ CC67,  1900, La Plata, Argentina
\\ Email: \email{harari@df.uba.ar, mollerac@venus.fisica.unlp.edu.ar, 
roulet@venus.fisica.unlp.edu.ar}}

\abstract{We study the effects of the galactic magnetic field on the 
ultra high energy cosmic ray propagation. We show that the
deflections of the cosmic ray trajectories can have many important
implications such as (de)magnification of the cosmic ray fluxes by
lensing effects (which can modify the spectrum of individual sources),
the formation of multiple images of a source or the existence of
regions of the sky to which the Earth is almost blind. The appearance of
image pairs is related to the existence of critical curves in the
magnification maps, which divide regions in the sky where the images
have opposite parities. The results are pictorially illustrated as
the stretching and folding of a `sheet' describing the sky seen on
Earth. Making use of the most energetic AGASA events we emphasize the
need to know the cosmic ray composition and the structure of the
magnetic field when attempting to do detailed cosmic ray astronomy. 
}

\preprint{.}

\begin{document}

\section{Introduction}

Ultra high energy cosmic rays (UHECRs) beyond the ankle of the spectrum,
i.e. with energies above 5 EeV (1 EeV = $10^{18}$ eV), are most
probably protons or nuclei of extra-galactic origin. The reason is that
at those energies large rigidities would not allow their confinement
by the galactic magnetic field and would also lead to excessively
large anisotropies, added to the lack of plausible galactic sources
for their acceleration to such enormous energies. 

The major goal of existing and planned large air shower
detectors is to unravel the nature and origin of these UHECRs, and one
of the difficulties already encountered is the lack of obvious sources
near the arrival directions of the highest energy events observed.
Possible sources for these particles are AGNs, radio galaxies or
$\gamma$ ray bursts. At the highest energies observed ($ E \gsim$ 70
EeV) the attenuation of the fluxes by interactions with the cosmic
microwave background radiation requires that the sources be not too
far away, i.e. at less than $\sim 100$~Mpc in the case of protons
\cite{GZK} and even closer for heavier nuclei \cite{pu76}.

In their journey from their sources to the Earth, cosmic rays (CRs) are
deflected by magnetic fields. A regular intergalactic
field is strongly constrained ($B < 10^{-9}$G) by rotation measures of
extra-galactic radio sources. Galaxy clusters may have stronger
magnetic fields, coherent over Mpc scales. It has been shown 
\cite{gi80, si99} that 
a random component of strength $B \sim 0.1\ \mu$G of the magnetic field 
of clusters (e.g.  the Virgo cluster) can  lead to a difussive 
propagation of protons up to energies as high as 100 EeV as the cosmic 
rays exit or cross a cluster. Cluster magnetic fields can also lead to
a change in the arrival direction of the UHECR and to energy
dependent time delays, relevant for the detection of burst sources. 
They can also produce several images of a source \cite{si99}. The
magnetic field of our Galaxy ($B \sim$ few $\mu$G) can also lead to
sizeable deflections, and the study of these deflections for different
magnetic field models has been performed in refs. \cite{st97, me98},
stressing the need to correct the observed arrival directions at Earth
to obtain the direction of arrival outside the region of influence of
the galactic magnetic field. The magnetic field of the solar system
(few tens of $\mu$G) has negligible effects on the deflection of the
UHECR trajectories (although it may lead to some peculiar events when
CR nuclei disintegrate when interacting with solar photons 
\cite{geza}).

In this work we study in detail the effects of the galactic
magnetic field on the extra-galactic cosmic rays arriving to the
Earth. We will determine the CR deflections in different realistic
models for the magnetic field and for general cosmic ray composition
(i.e. charge $Z$). Applying this to the highest energy ($E > 40$ EeV)
AGASA events \cite{ag99} we illustrate the need to know the CR 
composition and to
understand the details of the magnetic field when trying to locate
the CR sources. 

We then study the formation of multiple images of a
source, something which turns out to be a quite common phenomenon.
We also show that the CR deflections produce a magnetic lensing
effect, which can sizeably amplify (or demagnify) the flux arriving
from any given source. Since this effect is energy dependent, 
the energy spectrum seen on Earth will be different from that of the
source. 

We also show that the magnification of the CR flux by the 
galactic magnetic field  becomes divergent for directions
along critical curves in the sky seen from the Earth (corresponding to
caustic curves in the `source plane', i.e. in the corresponding
directions outside the Galaxy). These caustics  move with energy
and as a caustic crosses a given source direction, pairs of
additional images of the source  appear or disappear. These and
other features of the multiple images will be addressed in Sections
\ref{images} and \ref{spectrum}. 
As a final remark, we will discuss the relation of these results
with the Liouville theorem, which implies that an isotropic flux outside
the Galaxy (something that is not expected at UHEs) should remain
isotropic on Earth.

\section{Cosmic ray propagation in the galactic magnetic field}

The trajectories of CRs arriving at Earth can be strongly affected by
the galactic magnetic field. A convenient way of computing the
trajectory of a nucleus of charge $Ze$ and energy $E$ arriving at
Earth from a given direction is to follow the trajectory of a particle
with the same energy and opposite charge that leaves the Earth in that
direction up to a point where the effect of the 
galactic magnetic field becomes negligible \cite{th68}. 
Cosmic rays of energy $E$ arriving from
extra-galactic space in the direction in which the antiparticle left 
the galactic
halo will arrive at Earth in the direction in which the antiparticle
left the Earth. The antiparticle trajectories leaving the Earth can be
obtained numerically solving the equation
\begin{equation}
\frac{d^2\vec{x}}{dt^2}=-\frac{Z e c}{E} \frac{d\vec{x}}{dt} \times
\vec{B}(\vec{x}), \label{mforce}
\end{equation}
using a realistic model of the galactic magnetic field.

\subsection{Galactic magnetic field models}

The Milky way, as well as other spiral galaxies, is known to have a
large scale regular magnetic field. This is mainly determined by
measurements of the rotation of the plane of polarization of radiation from 
pulsars and extra-galactic radio sources due to the Faraday effect. 
The rotation angle is proportional to the
integral of  the product of the electron density and the magnetic
field component along the line of sight. Although there are still
large uncertainties in these determinations and there is not a general
agreement about the magnetic field structure, some characteristics are
known \cite{be96}. The magnetic lines are thought to follow the
spiral arms. The local value in the Sun's vicinity is $B \simeq
2\ \mu$G and it points in the direction $\ell \simeq 80^{\circ}$.

Magnetic fields of spiral galaxies are distinguished between bisymmetric
(BSS) and axisymmetric (ASS). BSS (ASS) fields are antisymmetric
(symmetric) with respect to a $\pi$ rotation. In other words, in BSS
models the field reverses its sign between the arms (assuming that
there are two arms), while in ASS models it has the same
orientation. A further classification of magnetic structures
distinguishes fields that are symmetric (S) or antisymmetric (A) with
respect to the Galaxy's mid-plane \cite{be96}. In the S models, the
field has the same direction above and below the galactic plane, while
in A models it is reversed\footnote{Note that in \cite{st97,me98} the A 
and S denominations are interchanged.}. 

We use two combinations of the four classes of models to illustrate the
effects of the galactic magnetic field on the propagation of CRs. We
consider a bisymmetric model with even symmetry (BSS-S), which is the
preferred model for our Galaxy, and some of the results are shown also
for the axisymmetric model with odd
symmetry (ASS-A), which somehow represents a possible extreme
variation.  We adopt a structure and strength very similar to 
those used by Stanev \cite{st97}, but smoothed out in order to avoid
spurious effects due to discontinuities in the field or its
derivatives in the calculations. 

In the galactic plane ($z=0$) the field, directed along the spiral
arms, has strength 
\begin{equation}
B_{sp}=B_0(\rho) \cos\left(\theta - 
\beta \ln(\rho/\xi_0)\right),
\end{equation}
for the BSS configuration \cite{so83}. The angle 
$\theta$ is the azimuthal coordinate
around the galactic center (clockwise as seen from the north galactic
pole), $\rho$ is the galactocentric radial cylindrical coordinate,
$\xi_0=10.55$ kpc corresponds to the galactocentric distance of the
maximum of the field in our spiral arm and $\beta = 1/\tan p=-5.67$,
where the pitch angle is taken as $p=-10^\circ$.

For the ASS model we adopt
\begin{equation}
B_{sp}=B_0(\rho) \cos^2\left(\theta - \beta 
\ln(\rho/\xi_0)\right).
\end{equation}
The radial and azimuthal components in the galactic plane are
given by
\begin{eqnarray}
B_\rho&=&B_{sp} \sin p,\\
B_\theta&=&B_{sp} \cos p.
\end{eqnarray}
We take 
\begin{equation}
B_0(\rho)=\frac{3 r_0}{\rho} \tanh^3(\rho/\rho_1)
~\mu\rm{G}, 
\end{equation}
with $r_0=8.5$~kpc the Sun's distance to the galactic center and
$\rho_1=2$~kpc. This function has the $1/\rho$ behavior at $\rho >
4$~kpc as in \cite{st97} and \cite{so83} and goes smoothly to zero at the
galactic center \cite{so83}.

For the dependence on $z$, we consider a contribution coming from the
disk and another from the halo. For the symmetric S models
\begin{equation}
\vec{B}_S (\rho,\theta,z)=\vec{B}(\rho,\theta,z=0) 
\left(\frac{1}{2\cosh(z/z_1)}+
\frac{1}{2\cosh(z/z_2)}\right)
\end{equation}
with $z_1=0.3$~kpc and $z_2=4$~kpc. These scale heights are chosen so
that this smoothed model gives a good fit to the model used by Stanev
\cite{st97}. The field reversal at the galactic
plane in the antisymmetric A models is taken into account adopting
$B_A (\rho,\theta,z)=B_S (\rho,\theta,z) \tanh(z/z_3)$, with a
negligible scale height taken as $z_3=20$~ pc. 

\subsection{Cosmic ray deflections}

The gyroradius of a CR nucleus with $E/Z=10$~EeV  in a 
uniform magnetic field of strength 3~$\mu$G is slightly 
larger than 3~kpc. 
The typical value of the large scale galactic magnetic 
field is a few $\mu$G, and is approximately uniform over scales 
of the order of a few kpc. Thus  the
motion in the galactic magnetic field of nuclei with 
ratio $E/Z$ above 10~EeV  
should be in general not very different from a quasi-rectilinear
trajectory, with deflections away from the
straight path smaller as larger is the CR energy. 
At lower energies the CR path lengths in the galactic
magnetic field significantly increase with respect to
the case of rectilinear propagation \cite{ka72}.
A turbulent component in the galactic magnetic field 
adds diffusion to the CR motion, but this effect is
not expected to be significant at energies such that
$E/Z \gsim 1$~EeV \cite{lee95}. 

Figure 1 depicts  trajectories
of  nuclei with $E/Z=1$ and 10 EeV in
the BSS-S galactic magnetic field model, that we will take as 
the reference one to illustrate most of the effects discussed in 
this work. 
When $E/Z=1$~EeV the nuclei follow helical trajectories
that make a very large number of turns before their arrival to
the Earth. 
These highly twisted paths are not exclusive of finely tuned, 
isolated incoming directions, but rather are a typical feature
at these energies. Notice that the direction
of entrance to the Galaxy need not be close to the disk in order 
for the UHECR to get trapped along the spiral structure of
the magnetic field (dashed lines). 
When $E/Z=10$~EeV the CRs follow instead much straighter paths.

\FIGURE{\epsfig{file=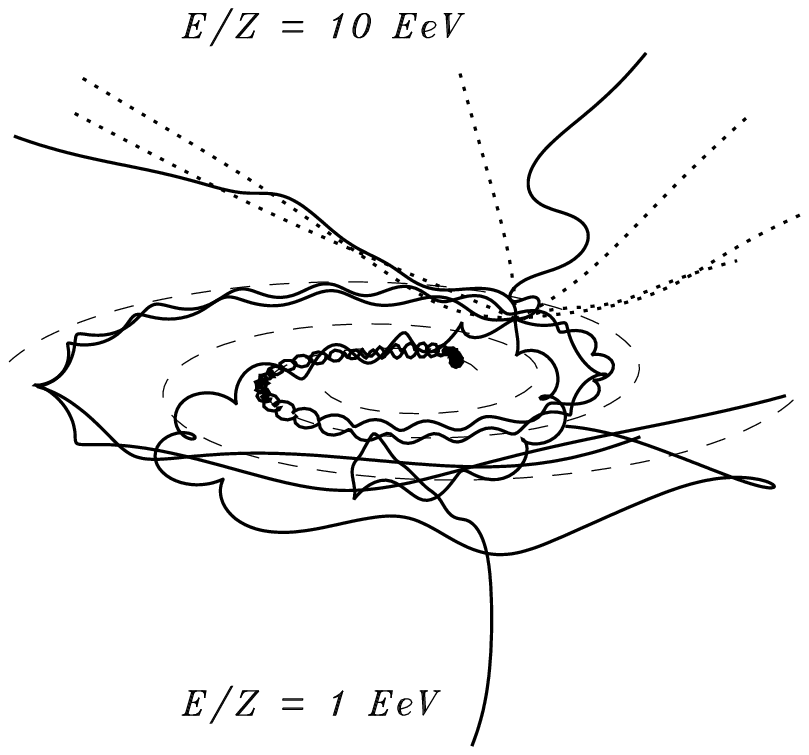,width=15cm}\caption{Examples of trajectories 
of nuclei with $E/Z=1$~EeV (solid lines) and $10$~EeV (dotted lines)
in the BSS-S galactic magnetic field model. Particles reach
the Earth at angles $\ell=30^\circ+n60^\circ$, and $b=-20^\circ$ for
$E/Z=1$~EeV and $b=20^\circ$ for $E/Z=10$~EeV.}}

The transition between  quasi-rectilinear and drift 
motion, in which CR trajectories turn around several times before 
they reach the Earth, occurs rather sharply at values below 
$E/Z \sim  3$~EeV.
Indeed, an average over a regular grid of arrival directions
of the ratio between the distance traversed by a CR 
within 20 kpc of the galactic center and the distance it would have
traversed along a straight path gives a mean value of 1.01, 1.14 
and 2.2 for ratios $E/Z=$ 10, 3 and 1 EeV respectively (in the
reference BSS-S magnetic field model).
The rms dispersions are 0.01, 0.18 and 1.7 respectively.
Let us note that most of the excess corresponds to the increased path
length within $\sim \pm 1$~kpc of the galactic plane.

Figure 2 illustrates how much do the arrival directions
of  charged UHECRs deviate from their incoming direction at the
galactic halo due to deflections in the magnetic field.

\FIGURE{\epsfig{file=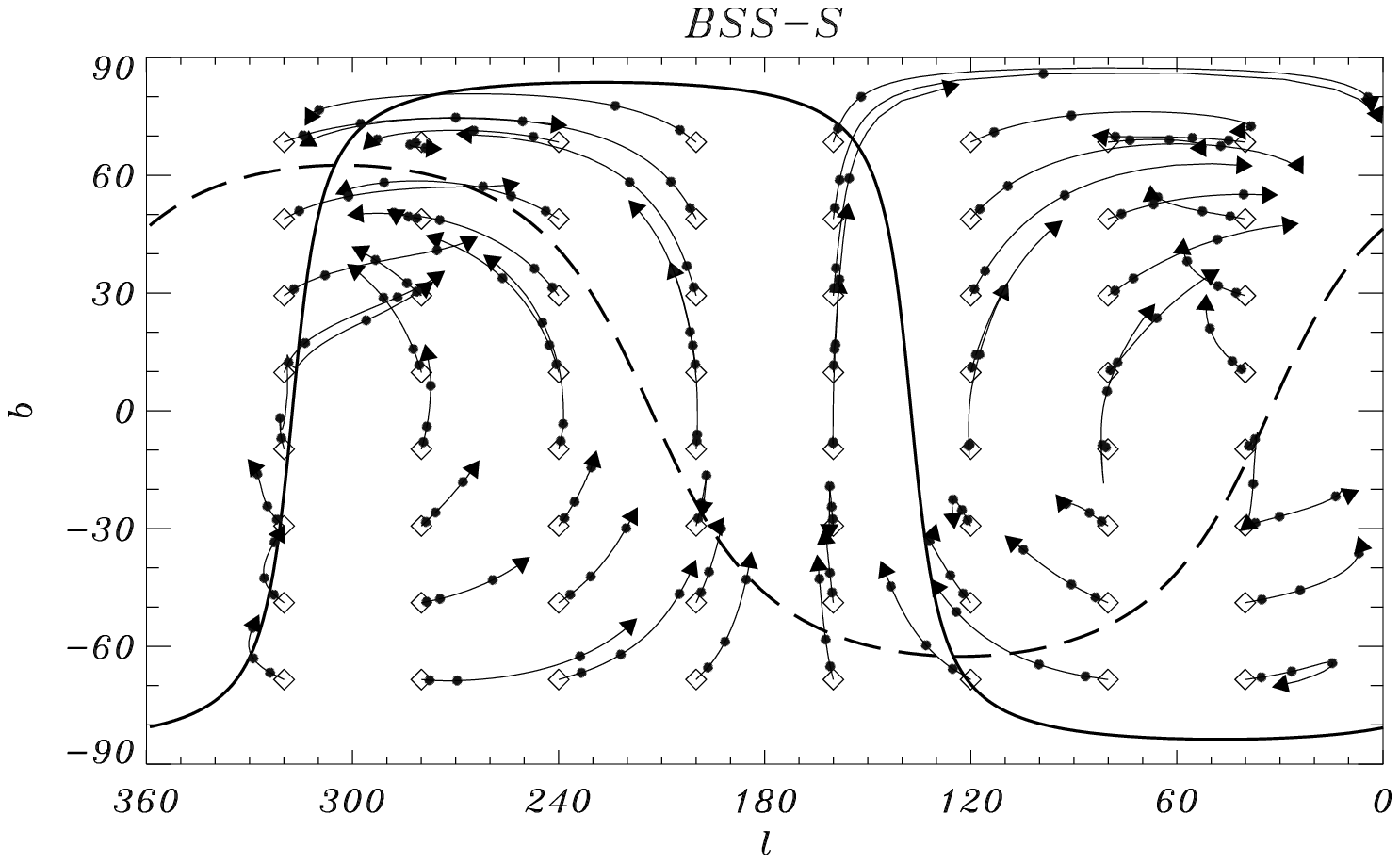,width=12.5 cm}
\epsfig{file=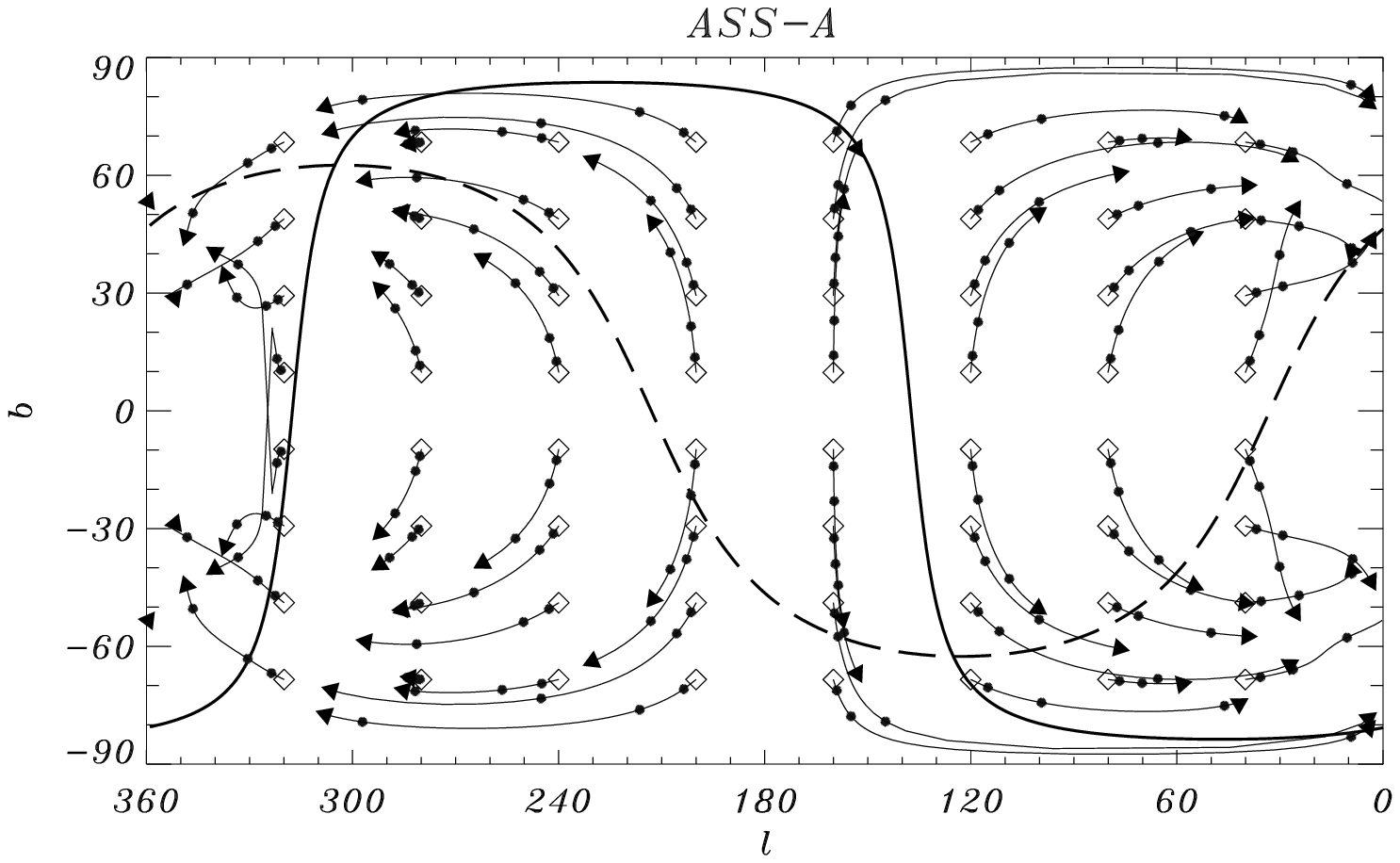,width=12.5 cm}\caption{
Deflections of CR nuclei in the galactic
magnetic field. Each diamond is an  arrival direction at Earth.
The points along each line as we move away from the diamond 
are actual incoming directions at 
the halo for decreasing CR energies. 
The dots correspond to $E/Z=$ 100, 30 and 10 EeV 
respectively. The tip of each arrow corresponds to $E/Z=7$~EeV.
Top panel: BSS-S magnetic field configuration. Bottom panel: 
ASS-A model.}}

The plot is in galactic coordinates $\ell$ (longitude) and 
$b$ (latitude). The super-galactic plane (solid line) and the 
Earth equatorial plane (dashed line) are superimposed for visual 
guidance. A regular grid of UHECR arrival directions at 
Earth is mapped into the directions from which 
CRs of different energies entered the halo.
In practice, the trajectories are followed up to distances
20 kpc away from the galactic center, where the effects of the
magnetic field in the models considered here are already negligible.
The diamond at the origin of each line denotes the observed 
arrival direction.
The points along the line from each diamond towards the tip of the 
respective arrow indicate the actual direction from which the CR 
entered the galactic halo for decreasing CR energy. 
The dots along each line correspond to 
nuclei with ratio $E/Z=100$, $30$
and $10$ EeV respectively. The tip of the arrow corresponds to
the incoming direction at the halo of a CR with $E/Z=7$ EeV. 
The top panel displays the result for  the
BSS-S galactic magnetic field model and the bottom panel
corresponds to the ASS-A configuration.

This figure is the analogue of Figure 2 in Stanev's work \cite{st97}
\footnote{Note that in \cite{st97} $\ell$ actually
grows from left to right.}.
Deflections in the ASS-A model are symmetric with respect to the 
galactic plane, since $\vec B(z)=-\vec B(-z)$.
Deflections in the BSS-S case are neither symmetric nor
antisymmetric with respect to the galactic plane,
although there is an approximate antisymmetry at 
relatively high energies. 
In the ASS-A model there is a flow of lines out from the
galactic plane. CRs from extra-galactic  sources at relatively 
high latitudes will be observed at arrival directions
closer to the galactic plane. From the opposite viewpoint, this
flow of lines out from the galactic plane suggests that
CRs from extra-galactic sources that lie very close to the 
galactic plane may be practically unable to reach the Earth below 
certain energies. This will be discussed in the next Section.

Notice also in Figure 2 that there are directions in the sky where
the tips of the arrows converge, or even  where different lines
intersect.  This indicates the 
possibility of observing, for some range of energies, multiple 
images of a single  CR source, since one
incoming direction at the halo is mapped into two or more 
arrival directions at the Earth. This possibility will also
be discussed
at length in the next Section. There are other
peculiar features in Figure 2, such as sudden changes in
the direction of some arrows. The source of this behavior will
also become apparent from the discussion in the next Section.

As a concrete illustration of the impact that deflections 
in the galactic magnetic field may have upon  UHECR observations, 
we plot in Figure 3 the arrival directions of all the 
events recorded by AGASA \cite{ag99}
with energies higher than 40 EeV and the corresponding
incoming direction at the galactic halo, depending upon the assumed CR 
composition. 
On top  are the 17 AGASA events with energies higher
than 60 EeV, and at the bottom are the 30 events with energies between 
40 and 60 EeV (plus one event with energy just below 40 EeV, also 
included in Table 2 in \cite{ag99}).  
The diamonds denote the observed arrival direction of each event.
The points along each line denote the actual incoming direction
at the halo for nuclei with increasing $Z$ as we move away from each
diamond towards  the tip of the arrow. In the top panel the dots
along each line indicate the results for protons ($Z=1$), Carbon ($Z=6$), 
Neon ($Z=10$), Silicon ($Z=14$), Calcium ($Z=20$), and the tip of each 
arrow corresponds to Iron ($Z=26$). In the case of the lower energy
events $Z$ ranges from 1 to 10. The deflections shown correspond to
the BSS-S magnetic field configuration.

\FIGURE{\epsfig{file=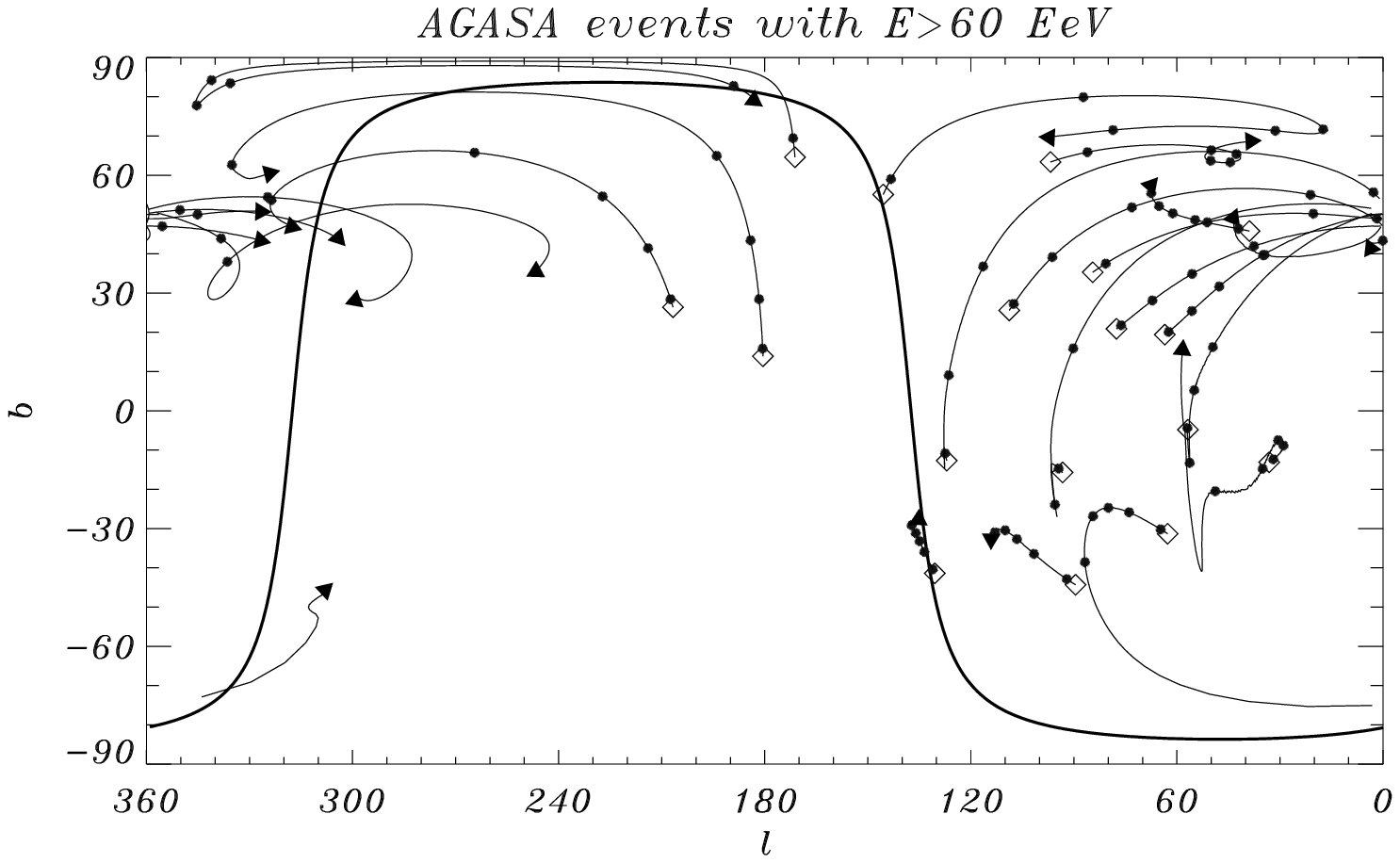,width=12.5truecm}
\epsfig{file=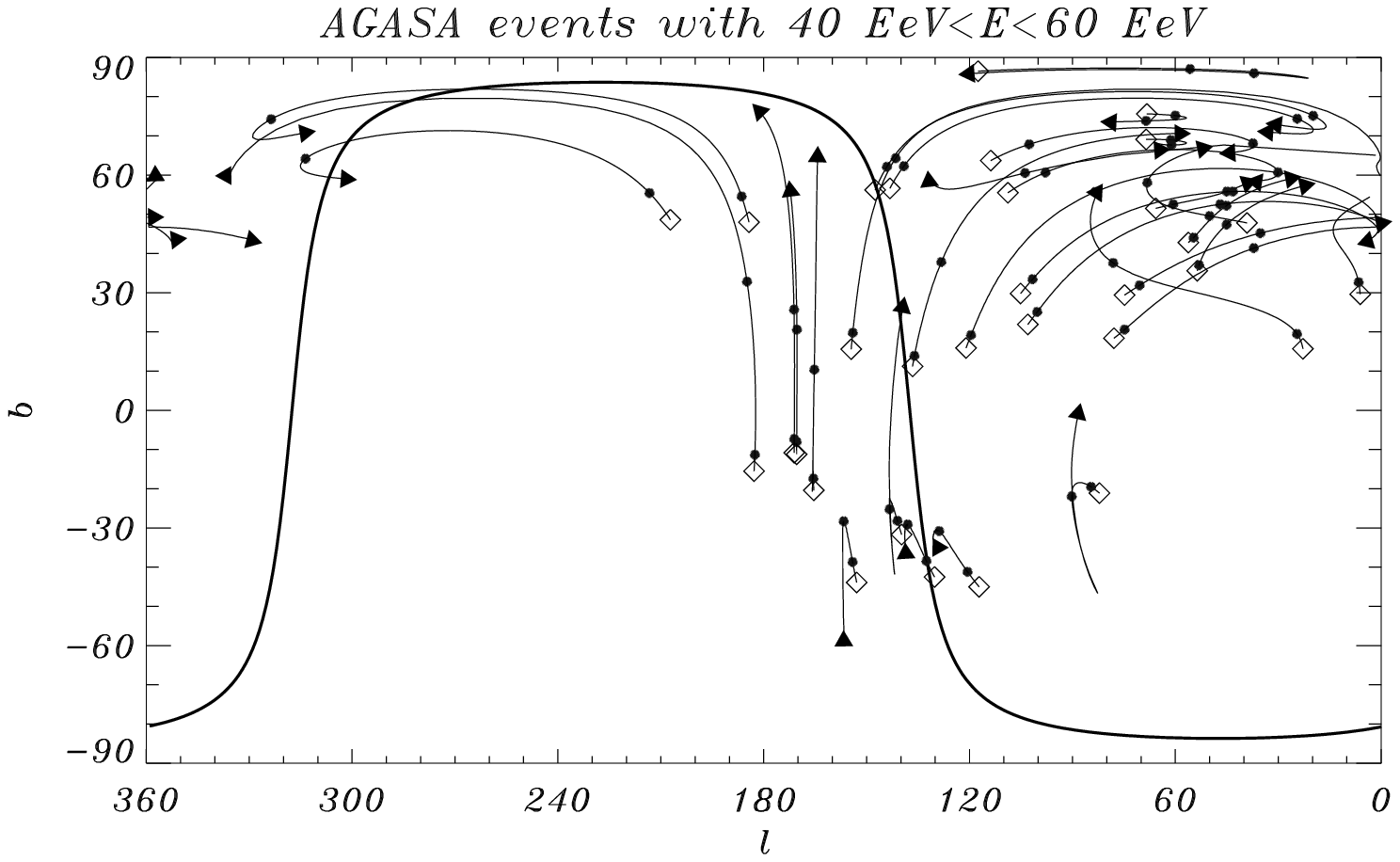,width=12.5truecm}\caption{
Observed arrival directions (diamonds) of AGASA events 
with energies larger than 60 EeV (top panel) and energies
between 40 EeV and 60 EeV (bottom panel), and the corresponding
incoming directions of CRs at the galactic halo for increasing $Z$ as we
move away from each diamond. The dots indicate the results for $Z$=1, 6, 
10, 14 and 20 and the tip of the arrow is for $Z=26$ (Iron)
in the top figure. In the bottom figure $Z$ varies between 1 and 10
only. Deflections correspond to the BSS-S model.}}

Correlation between the observed events and known sources for a 
heavy (large $Z$) component of the UHECRs is clearly impossible without a
detailed  knowledge of the galactic magnetic field structure.
Besides, clustering of observed arrival directions
may not reflect clustering of the incoming directions at the
galactic halo. The opposite is also a possibility:  several quite 
separated events may arrive from the same or nearby incoming 
directions at the halo for particular values of $Z$. 

\section{Multiple images of cosmic ray sources}\label{images}

A very interesting and quite common effect resulting from the magnetic
deflections is the appearance of
multiple images of an extra-galactic UHECR source. This means that CRs
arriving to the galactic halo from a given direction can reach the
Earth following several different trajectories and are hence observed with
different arrival directions. This can clearly be appreciated in
Figures 4 and 5 where we have plotted for a regular grid of arrival
directions at Earth in galactic coordinates $b$ and $\ell$, the
direction from which the particles arrived to the galactic halo. 

We move along `horizontal' or `vertical' lines following  
arrival directions at
Earth with fixed $b$ or $\ell$ respectively.  
As the trajectories depend on the
$E/Z$ ratio, the surface is stretched and folded as $E/Z$ changes.
The three panels
in Figure 4 show the development of the folding for the BSS-S model
at energies $E/Z=$ 30, 10 and 5~EeV.
For large $E/Z$, the sheet is quite smooth as the effect of
the  magnetic field on the CR trajectories becomes small. For
decreasing $E/Z$, foldings on the surface appear. 
CRs arriving from a direction in the halo where a folding has
developed 
are seen at Earth from all the points of the grid that overlap in that
place. For a source located in a direction in
which a folding develops at small energies, we observe that the
high energy apparent position of the source moves gradually for
decreasing energies and at the energy at which the folding reaches the
source direction, a pair of new images of the source appears in a
different region of the sky. Alternatively, if a folding moves out of
the source direction, a pair of images merges and disappears (if the
sheet were to unfold just along the direction of the source, the three
images would merge together into a single one, which is the magnetic
analogue of crossing a caustic through a cusp in gravitational lensing
theory). 
Notice for example that if a region which is folded in one
direction (leading to a piece of sky with three images) develops another
folding in a transverse direction as the energy decreases, six new
images of a source located there will appear (leading then to a total
of nine images for that source). 
Since it is apparent from the Figures that large areas of the sky
develop foldings, multiple images should be a pretty common
phenomenon. The fact that new images appear in pairs is a known fact
in gravitational lensing theory \cite{sch92}.

Another interesting fact that can be appreciated in Figures~4 and 5
regards the parity of the images observed. When a folding in the
surface develops and a couple of new images of some source appears,
one of the new images is a mirror reflected image of the source. In
general every folding separates regions of images with opposite
parity. 

\FIGURE{\epsfig{file=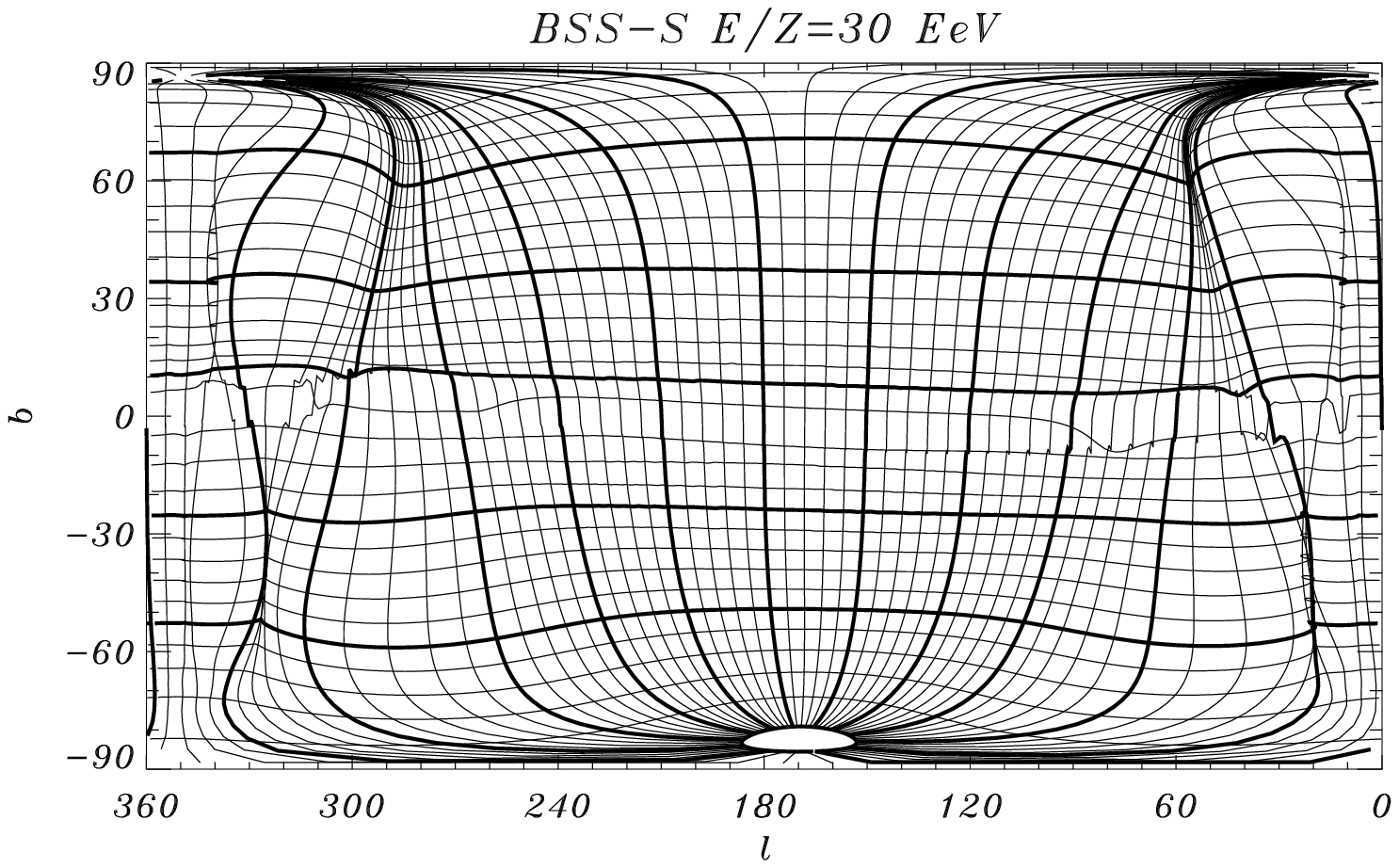,width=12cm}
\epsfig{file=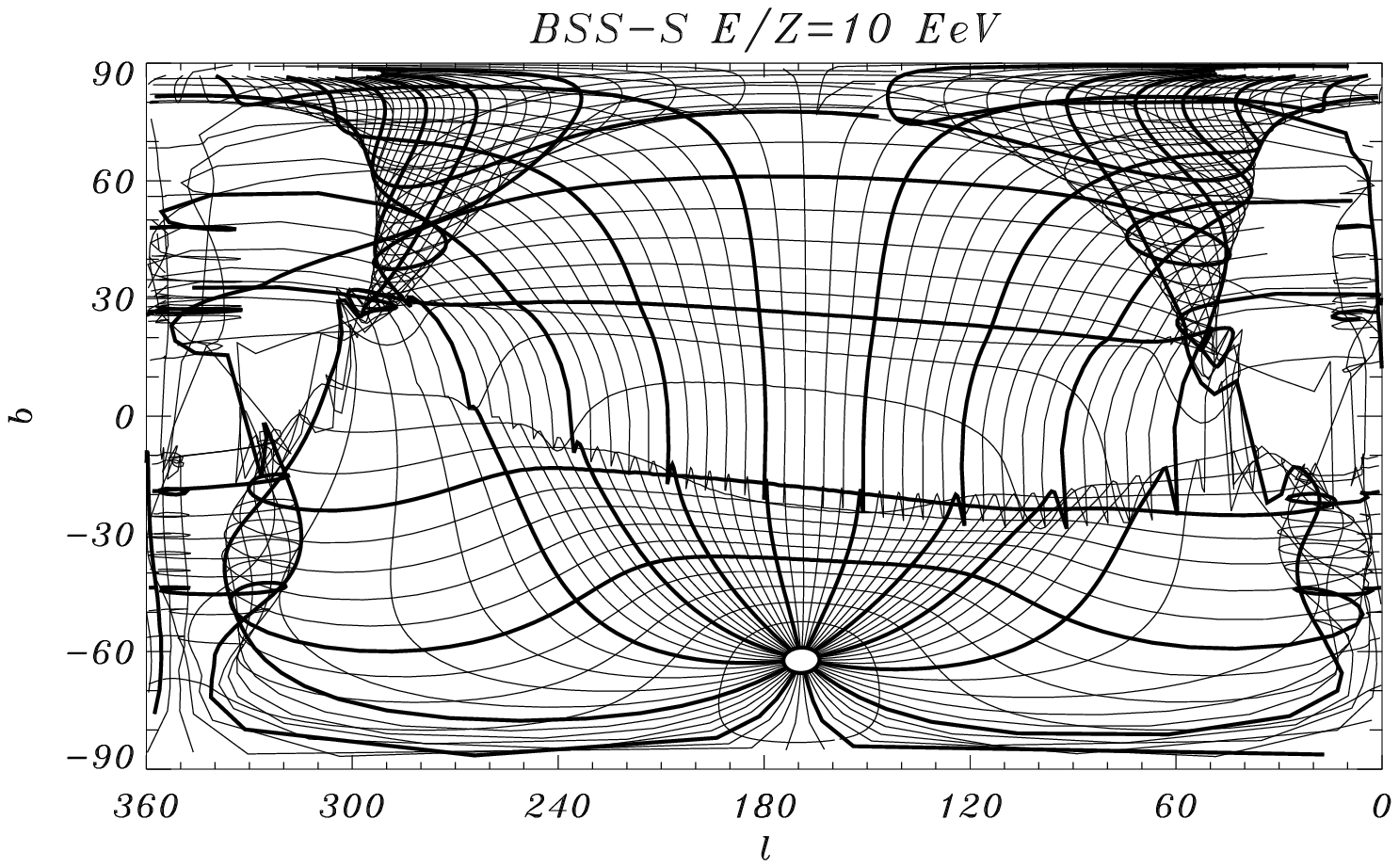,width=12cm}
\epsfig{file=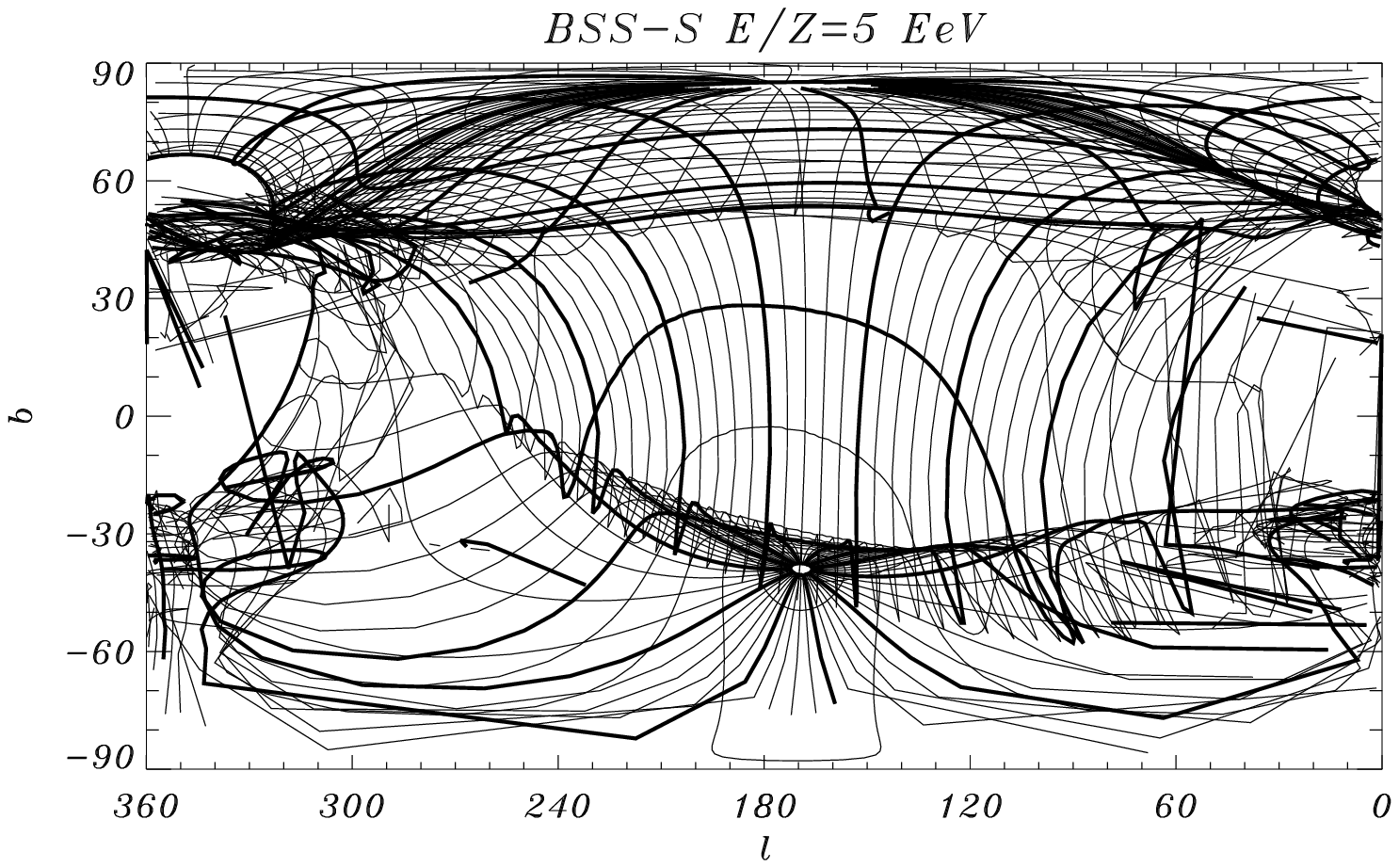,width=12cm}\caption{``Sky sheets'':
directions  
of incoming CRs in the halo that correspond to a regular grid
of arrival directions at Earth, for the BSS-S magnetic field 
configuration with $E/Z=$ 30 (top), 10 (middle) and 5 EeV (bottom).
Sources located in regions where the sheet is folded have multiple
images.}}

\clearpage

\FIGURE{\epsfig{file=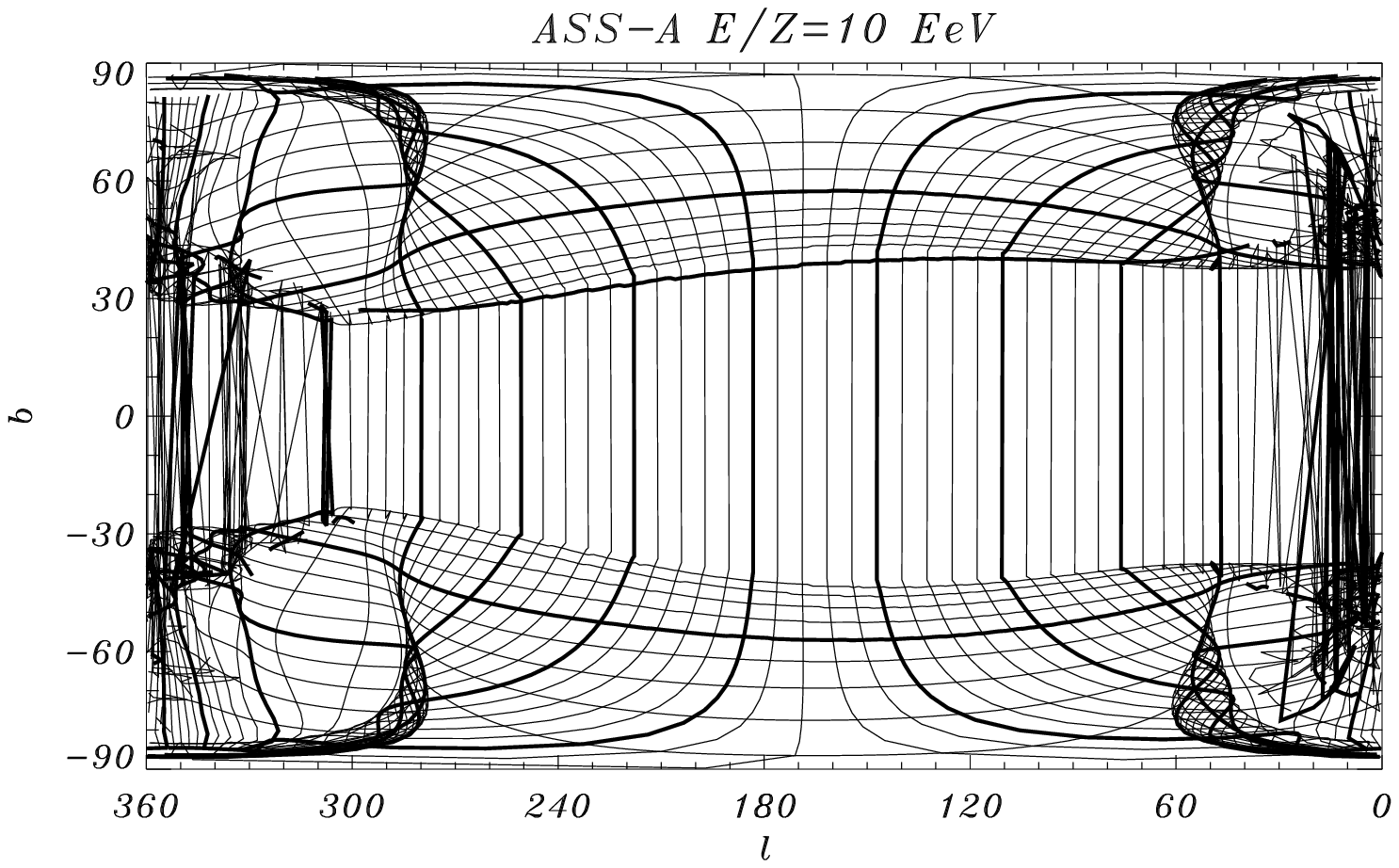,width=12cm}\caption{Same as Fig. 4
for the ASS-A magnetic field configuration and for $E/Z=10$~EeV only.
The Earth is practically blind to CR sources located in the very
stretched region around the galactic plane.
}}

Another curious effect that manifests in Figures 4 and 5 is the 
appearance of regions where the sheet is very stretched, and hence the
incident fluxes will appear very demagnified. This means that CR 
sources located in those
regions will become almost undetectable from the Earth. 
This is a rare effect in symmetric S
models, associated essentially to regions near the galactic
center. However, for the antisymmetric A models it is a very common
effect, as the sky sheet presents a highly demagnified region all 
along the galactic plane,
as shown in Figure~5. The reason is that antiparticles leaving the
Earth with small angles above or below the galactic plane are
deflected away from the plane. These deflections increase with
decreasing energies and are already larger than $30^\circ$ for
$E/Z=10$~EeV if $|b|>0.1^\circ$ in the Earth.
Thus, for the sources located in a large area of the sky
around the galactic plane ($|b|< 30^\circ$), the CRs can reach
the Earth only if they come along a fine-tuned path which arrives very
close to the galactic equator. We note that if we were to take $z_3\to 0$
(so that the change in sign of $B$ across the galactic plane 
in the A models were discontinuous), there would be no paths
reaching the Earth from sources within $\pm 30^\circ$ in Figure
5. Hence, for all practical purposes it would be as if the ``sky sheet''
were teared up and divided into two disconnected pieces. These kind of
tears appear when there is a discontinuity, i.e. when infinitesimally
nearby trajectories leaving the Earth lead to widely separated
directions in the halo\footnote{This is also the reason why in
microlensing only two images exist when a point-like lens is
considered. The third one falls in a tear with zero magnification.}.
In the S models (Figure~4), something qualitatively similar happens
near the galactic center, and will lead to the ``loss'' of images in
highly demagnified regions (see below).

{} From the data used to construct the grids in Figure 4 one can read 
the observed angular position(s) at a given energy of the image(s) 
of a source that injects CRs at a fixed direction in the halo. 
One can then follow the displacement in the observed angular position of 
each image as the CR energy is gradually changed, exploring at each 
energy step a neighborhood of the arrival direction from the
previous step, and backtracking it to the halo until it reaches the 
appropriate injection direction with the required accuracy.  
One should start this process at low energies and then move up,
otherwise one would only follow the evolution of the principal image 
of the source but miss the secondary images.

\FIGURE{\epsfig{file=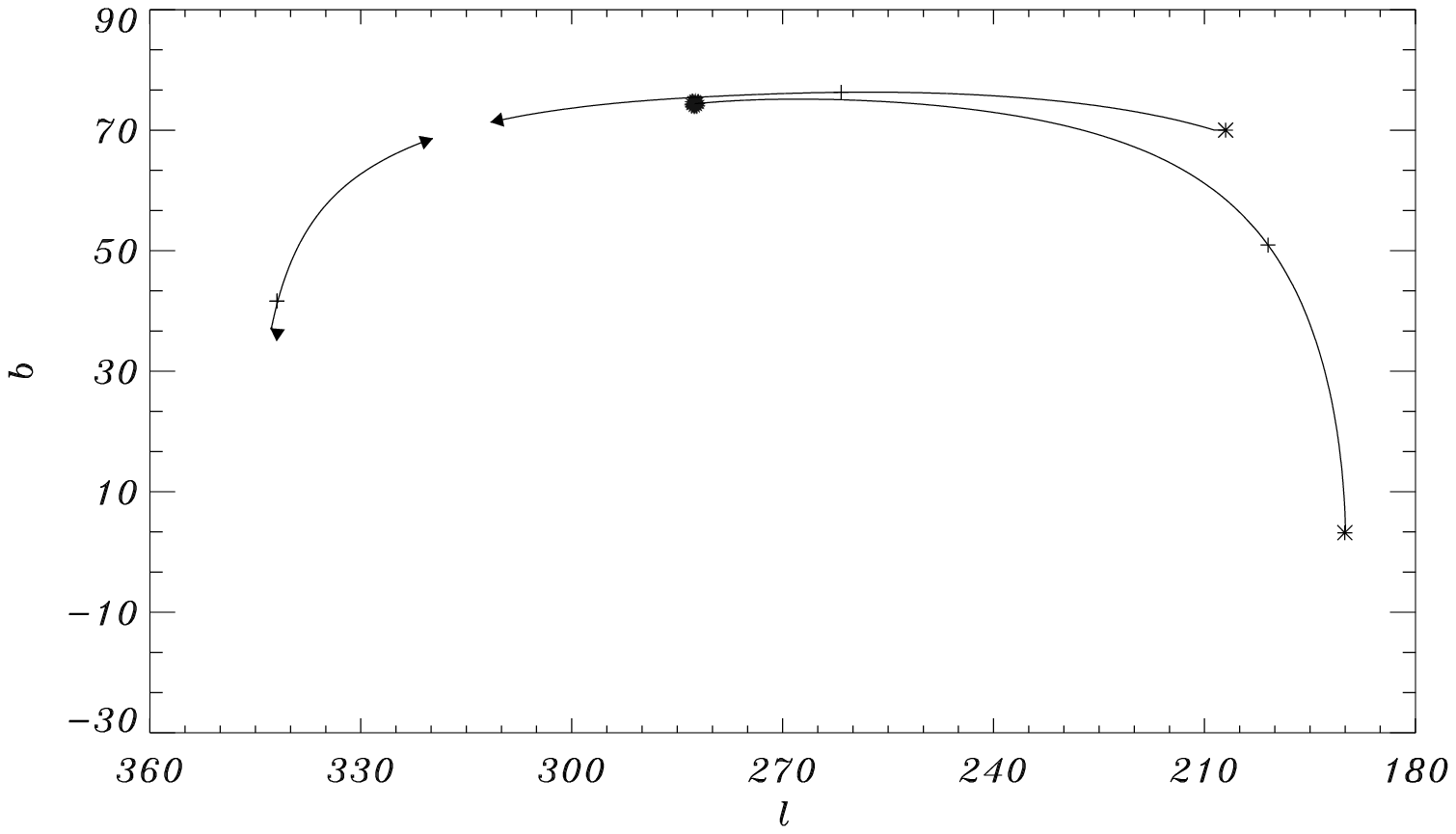,width=11 cm}
\epsfig{file=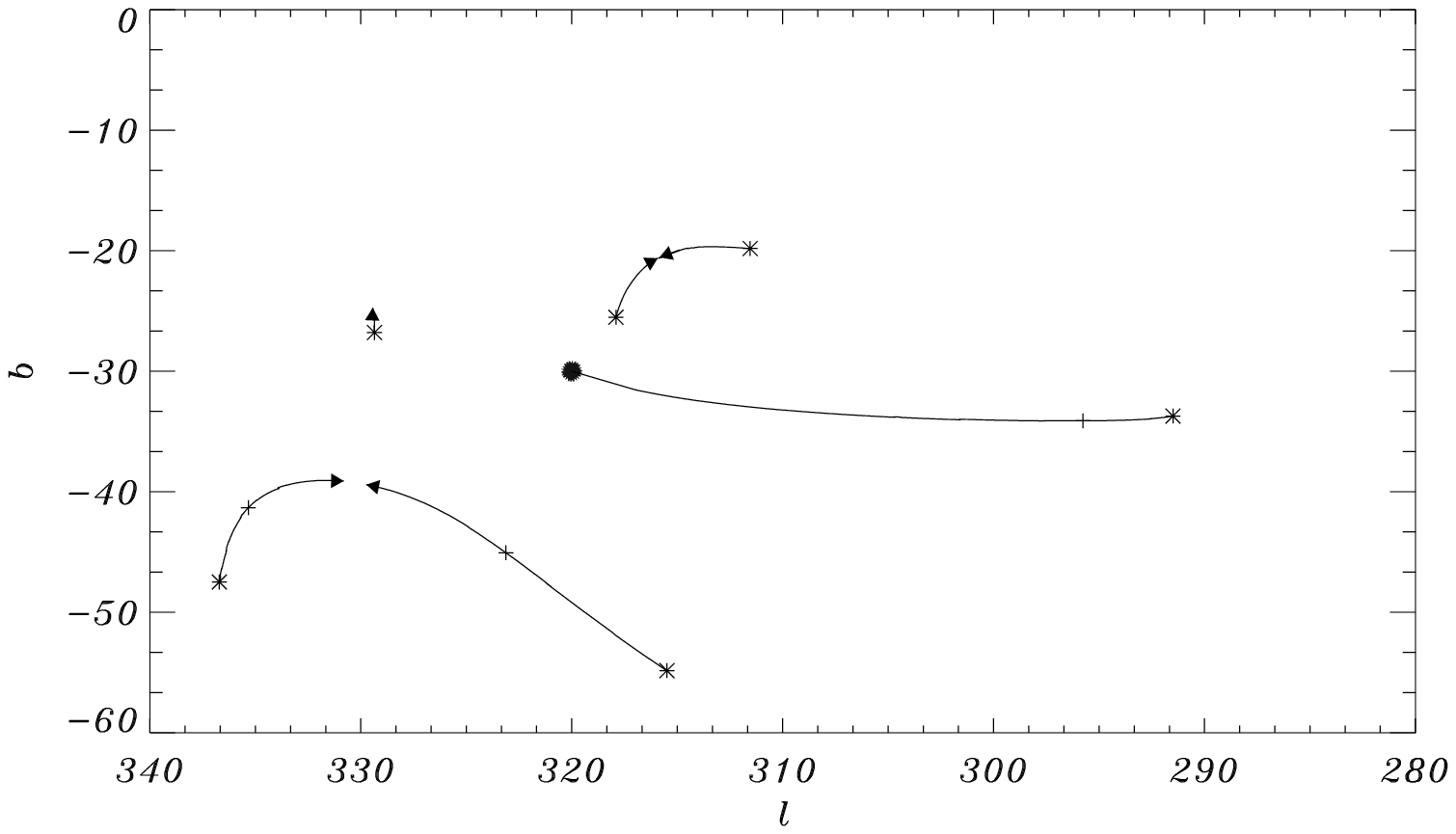,width=11 cm}\caption{Illustration of
the angular displacement of a source image as a function of energy, 
and of the formation and disapearance of secondary images. The actual
source in the top panel is at $(\ell,b)=(284^\circ,75^\circ)$ (M87) 
and the source in the bottom panel is at $(\ell,b)=(320^\circ,-30^\circ)$.
The magnetic field is taken in the BSS-S configuration.}}

Some examples of the effects discussed are illustrated in Figure 6
for two particular source directions, both lying near the
super-galactic plane, adopting the BSS-S magnetic field configuration. 
The top panel corresponds to M87
in the Virgo Cluster (galactic coordinates
$(\ell,b)=(284^\circ,75^\circ)$).  
The dot is the actual source direction, which is the
observed one at very high energies. As the $E/Z$ ratio decreases, the
apparent position moves along the line. When $E/Z$ reaches around
20~EeV, a pair of new images are created at the point where two
arrows meet, and the images then move along the lines shown, one to
the right and the other to the left. The plus signs indicate the
position of the images at $E/Z=10$~EeV. Just below this
value the image on the left falls into a region of negligible
magnification near the galactic center and hence 
``disappears'', in the sense that the path leading to it is so
fine-tuned that we were unable to find it. The asterisks show
the locations of the two surviving images at $E/Z=5$~EeV. Note that
the images become separated by more than 150$^\circ$ from each other.

The bottom panel in Figure 6 displays an example in the 
southern hemisphere,  for a
direction at $(\ell,b)=(320^\circ,-30^\circ)$. Here two images are
created at $E/Z\simeq 15$~EeV and another couple at $E/Z\simeq
6$~EeV. There is also an additional image which appears just above  
 $E/Z\simeq 5$~EeV (its companion image is very demagnified and we
couldn't find it). Three images are then observed at
$E/Z\simeq 10$~EeV (plus signs) and a total of six are `visible' 
at $E/Z\simeq 5$~EeV
(asterisks)\footnote{It is remarkable that these kind of `footprints'
are found in the sky when looking beyond the ankle of the CR spectrum.}.    

Finally, let us note that 
in gravitational lensing theory, the formation of multiple images can
also be studied using Fermat's principle, i.e. looking for the paths
between the source and the observer for which the light travel time is
an extremum. Similarly, each CR trajectory arriving to the Earth is
an extremum of the action for a charged particle in the galactic magnetic
field, i.e. of
\begin{equation}
S=\int_{Source}^{Earth} {\rm d}t \cal{L},
\end{equation}
where the Lagrangian is
\begin{equation}
{\cal L}=-m c^2 \sqrt{1-(v/c)^2} +\frac{q}{c} \vec{v}\cdot\vec{A},
\end{equation}
with $\vec{A}$ the magnetic field vector potential. The different extrema of
the action ($\delta S = 0$) will give rise to the different paths
along which the CRs can arrive to the Earth, and hence lead to the
different images of the source. 

We note that, since $|\vec{v}|$ is constant along the real
trajectories, the kinetic piece of the action is just proportional to
the path length traversed. At very high energies, $E/Z> 100$~EeV,
only the kinetic term is relevant in the functional variation of the
action, and hence a single image will be
seen, corresponding to the almost straight propagation. This path
is a minimum of the action.

For lower energies, the magnetic contribution becomes relevant and
competes with the kinetic piece to produce new extrema, which can 
in principle be saddle points (in the sense that there will be (not
infinitesimally) neighboring paths for which 
the action will increase, and others for which it will decrease) or
even maxima.

\section{The observed spectrum}\label{spectrum}

\subsection{Energy-dependent flux amplification}

As we mentioned before, the galactic magnetic field can act
 as a giant lens that
amplifies or demagnifies extra-galactic sources.
 The magnitude of this
effect depends on the direction of observation and on the $E/Z$
ratio. Consequently, the observed spectrum of a source can be
 affected,
 as the CR flux will suffer different amplifications for different
energies. In the case of multiple imaging of a source, each of the
images will be magnified by a different amount by the magnetic field.

This can be studied quantitatively by considering the focusing effect 
 of the galactic magnetic field 
on a bundle of particles arriving from outside the Galaxy.
When a bundle of charged particles with parallel trajectories arrives
to the galactic halo and, after being deflected by the galactic
magnetic field, reaches the Earth, it will be generally more focused
or defocused. The amplification is given by the ratio of the flux of
particles reaching the Earth 
 to that outside the galactic halo. We compute
the magnification of a source observed at Earth in a given direction
and for a given $E/Z$ ratio by first backtracking an antiparticle
 leaving in
that direction up to a distance of 20~kpc from the galactic center,
where the effect of the magnetic field becomes negligible. Then, we
track the particle back to the Earth by reversing the final velocity
of the antiparticle at the halo border. At the same time, we track
also the position of two more particles entering the halo with
trajectories parallel to the fiducial one, the same $E/Z$ value, but
displaced in two orthogonal directions. We follow the evolution of the
displacement of these two particles from the fiducial trajectory by
solving the equations
\begin{equation}
\frac{d^2\Delta\vec{x}_i}{dt^2}=
\frac{Z e c}{E} \left(\frac{d\vec{x}_0}{dt} \times
\Delta\vec{B}_i(\vec{x}_0)+\frac{d\Delta\vec{x}_i}{dt} \times
\vec{B}(\vec{x}_0)\right) ,\label{disp}
\end{equation}
coupled to eq. (\ref{mforce}). In the previous equation, $\vec{x}_0$
denotes the position of the fiducial particle, $\Delta\vec{x}_i$ 
($i=1,2$) the displacements of the other two particles, and 
$\Delta\vec{B}_i(\vec{x}_0)\equiv \vec{B}(\vec{x}_0+\Delta
\vec{x}_i)-\vec{B}(\vec{x}_0) \simeq
(\Delta\vec{x}_i\cdot\nabla)\vec{B}|_{x=x_0}$.
Note that in eq. (\ref{disp}) we have neglected a term proportional to 
$\Delta\vec{x}_i \times \Delta\vec{B}_i$ as it is of second order in
small displacements. When the three particles arrive back to the Earth
we compute the area subtended by the two displacement vectors as
$A_E=(\Delta\vec{x}_1 \times \Delta\vec{x}_2)\cdot\hat{v}|_E$,
obtaining the magnification associated to that direction as the ratio of
the original area subtended at the halo border, $A_H$, and $A_E$. 

We note that the linearity of eq. (\ref{disp}) in $\Delta \vec{x}$
implies that, if $\Delta \vec{x}(t)$ is a solution, so is $\alpha\Delta
\vec{x}(t)$, with $\alpha$ a constant. This implies
that all the points inside the initial rectangular area $A_H$ 
remain inside a parallelogram as they propagate, and arrive inside the
$A_E$ defined above.

A dimensional analysis may be helpful to estimate under which
conditions large (de)magnifications may be attained in a 
magnetic field. Assume that
the conditions are such that the deflection of the fiducial 
trajectory in eq. (\ref{disp}) is small. Take then  $\vec{x}_0\simeq s 
\hat s$, with $\hat s$ the unit vector along the 
initial direction
of the fiducial trajectory and $s$ the distance from the
entrance point at the halo border. 
The  deflection angle is thus given by 
$\delta\simeq (Ze/E)|\int_0^L \vec B_\perp (s)ds|
\simeq 5^\circ Z (10~{\rm EeV}/E)|\int_0^L \vec B_\perp (s)ds|/(1~
\mu{\rm G}\ 1~{\rm kpc})$. 
Here $\vec B_\perp$ is the component of $\vec B$ perpendicular
to the trajectory, and $L$ is the distance traversed by the
particle in the magnetic field. Assume also that the relative deflections
of two neighboring trajectories
initially parallel to the fiducial one
but displaced in orthogonal directions 
are also small, which implies small (de)magnifications. 
To the lowest order of approximation, the rhs of eq. (\ref{disp})
can be evaluated along undeflected trajectories, and 
the rate of change of the area $A$ subtended by the 
two displacement vectors is then given by
\begin{equation}
\frac{d^2A}{ds^2}\simeq -
\frac{Z e}{E} \hat s \cdot(\nabla\times\vec B)A_0\ ,\label{dA}
\end{equation}
where $A_0$ is the area initially subtended (and the initial condition
d$\Delta \vec{x}_i/$d$t=0$ was used).
After a distance $L$ the area becomes $A\simeq (1-2\kappa)A_0$ where
$\kappa =(Z e/2E)\int_0^L ds (L-s)\hat s\cdot
(\nabla\times\vec B)$.\footnote{The appearance of the factor $(L-s)$
here has a simple physical interpretation: the farther out that 
the lensing
effect induced by $\hat s\cdot\nabla\times\vec{B}$ takes place, the
more `leverage arm' it has to focus the beam into a smaller area.} 
The magnification is $\mu = A_0/A
\simeq 1+2\kappa$. We stress the fact that this conclusion applies
only to the case when both the deflection of the central
trajectory is small and $\kappa \ll 1$. Besides, changes in the shape
of the region subtended by the neighboring trajectories (its
potential distortion from a rectangle into an arbitrary parallelogram)
have also been neglected. In the language of gravitational
lensing, this result applies when both (de)magnification and
shear are small. 

Even though the previous analysis applies to the case of small
(de)magnification only, it serves its purpose to crudely estimate
the conditions necessary to attain large focusing,
which require $\kappa$ to be of order  unity.
Take, as a simple illustrative
example, an approximately straight trajectory perpendicular to the
galactic plane and incident from 
the north. In this case,  
$\kappa\simeq -0.9(Ze/2Er_0)\beta B_0(r_0)
\sin(\beta\ln(r_0/\xi_0))(z_1^2+z_2^2)\simeq 
0.13~(100~{\rm EeV}~Z/E)$, in the BSS-S model\footnote{Note that 
$\kappa$ will reverse sign for a CR entering from the south along the
vertical. Note also that for directions near the poles and small
magnifications, $\kappa\propto(z_1^2+z_2^2)$, and hence the halo field
gives the dominant contribution around the poles.}.
The corresponding deflection for the same incident trajectory is 
$\delta\simeq 0.8(Ze/E) B_0(r_0) \cos(\beta\ln(r_0/\xi_0)) (z_1+z_2)
\simeq 2^\circ~(100~{\rm EeV}~Z/E),$ and hence sizeable magnifications
do not require necessarily large deflections. 
For instance, we shall see that there are already divergent 
magnifications 
at $E/Z\simeq 30$~EeV, even along directions for which deflections are
not larger than $10^\circ$. 
Viceversa, a strong but
almost constant field can produce large deflections without
significant magnifications.  

\FIGURE{\epsfig{file=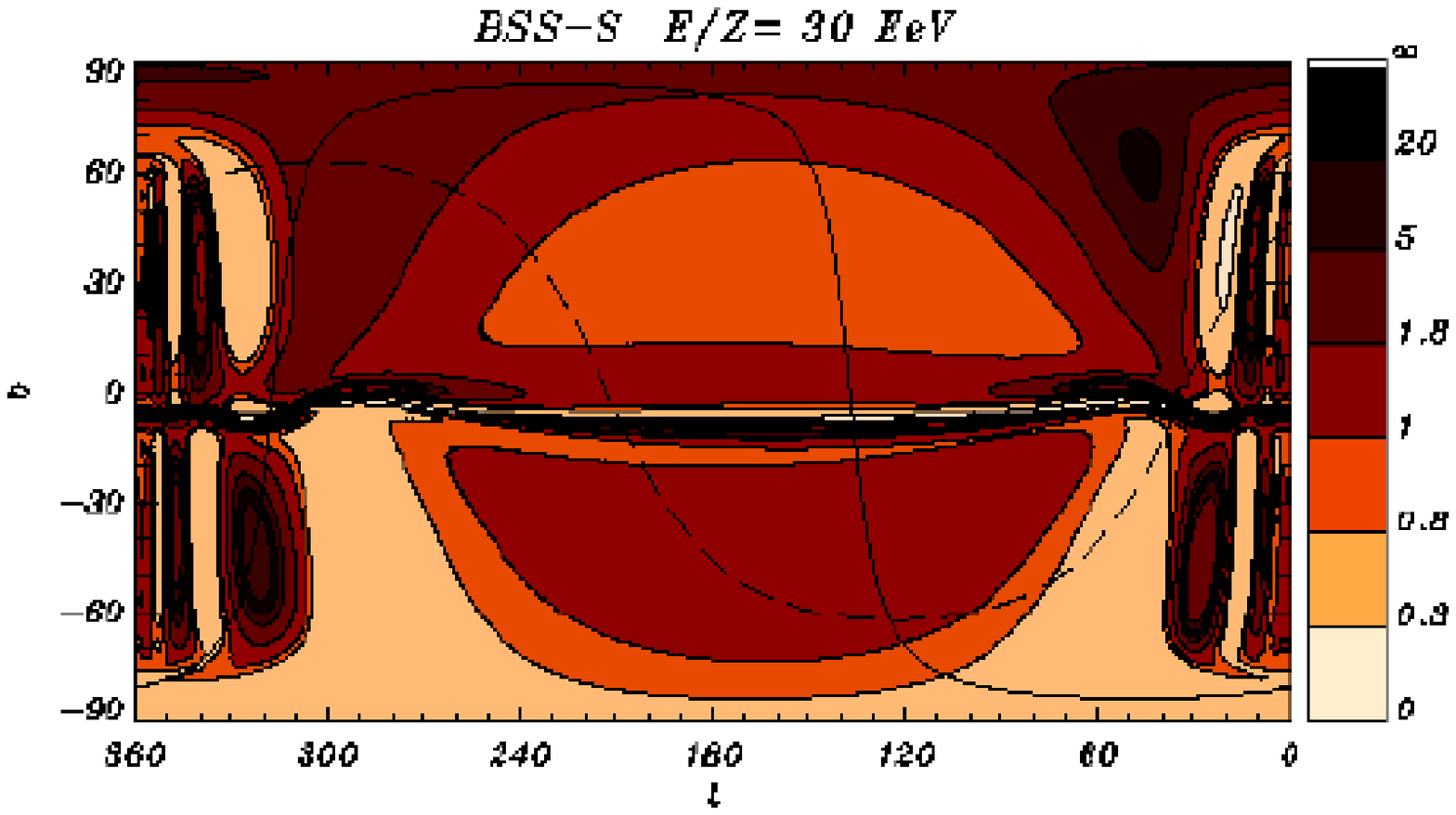,width=12cm}
\epsfig{file=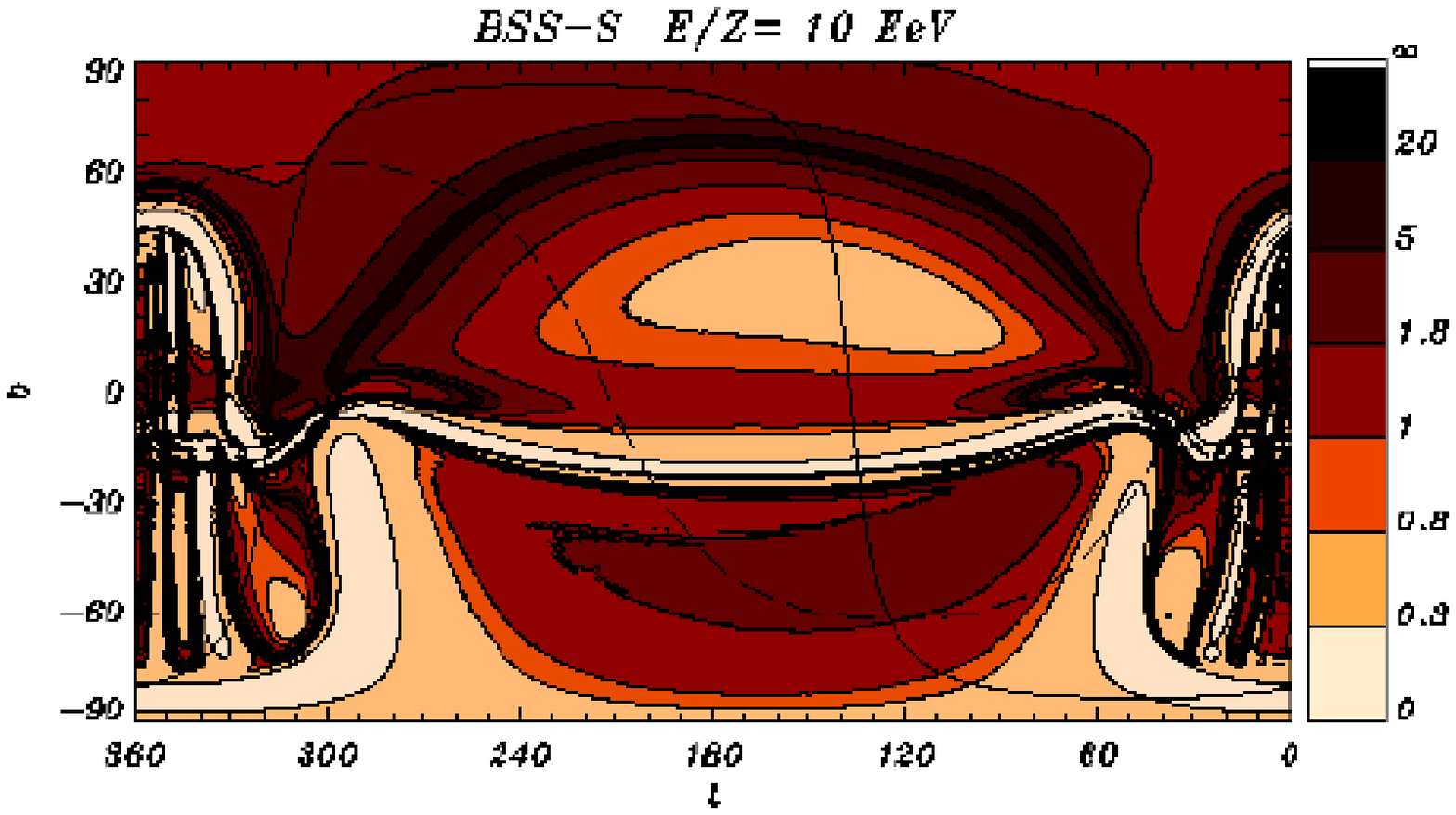,width=12cm}
\epsfig{file=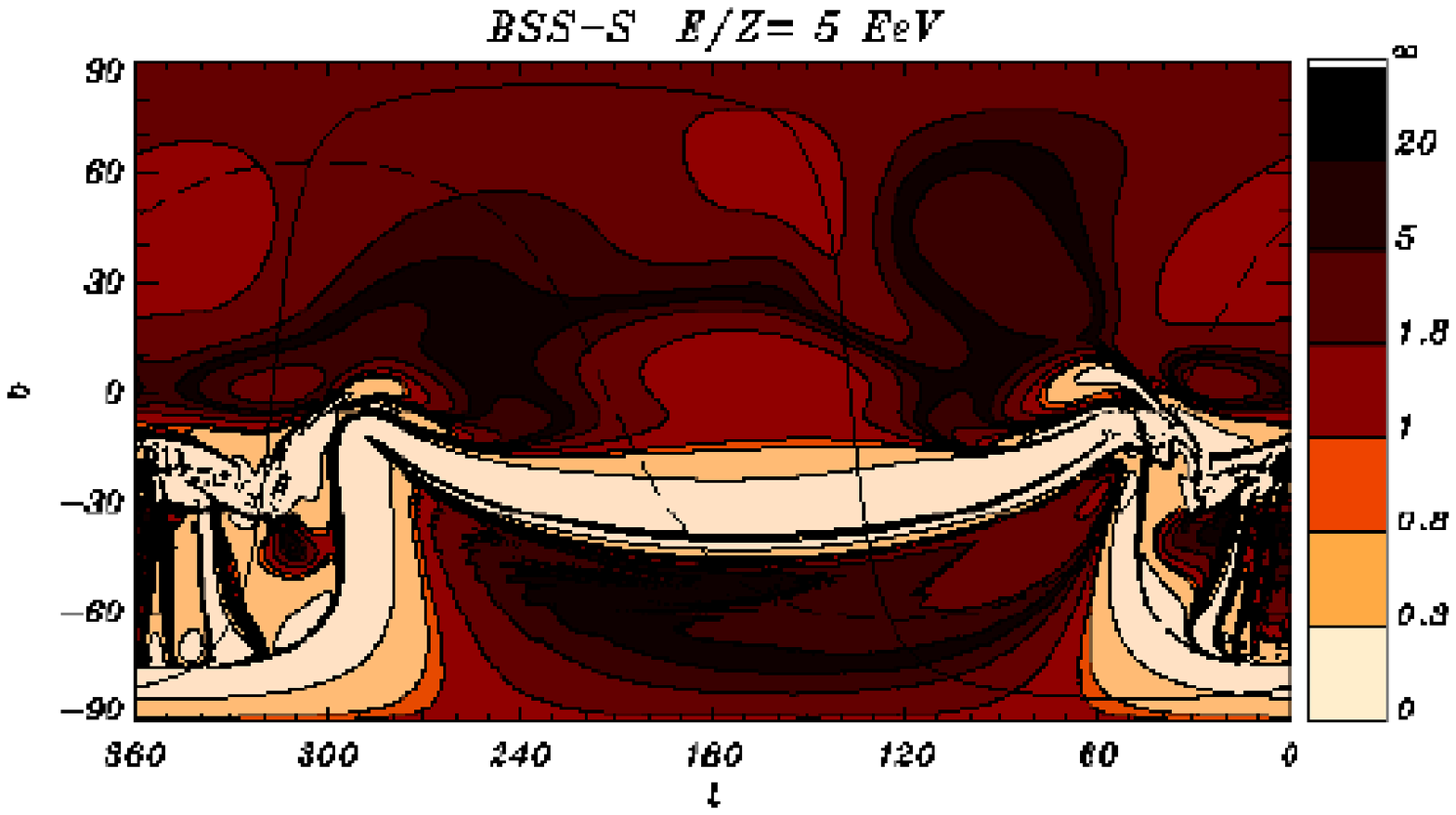,width=12cm}
\caption{Contour plots of
the magnification of the CR flux from a point source as a function of
the  arrival direction at Earth, in the BSS-S model and for
$E/Z=30$ (top), 10 (middle) and 5 EeV (bottom).}}

\clearpage

\FIGURE{\epsfig{file=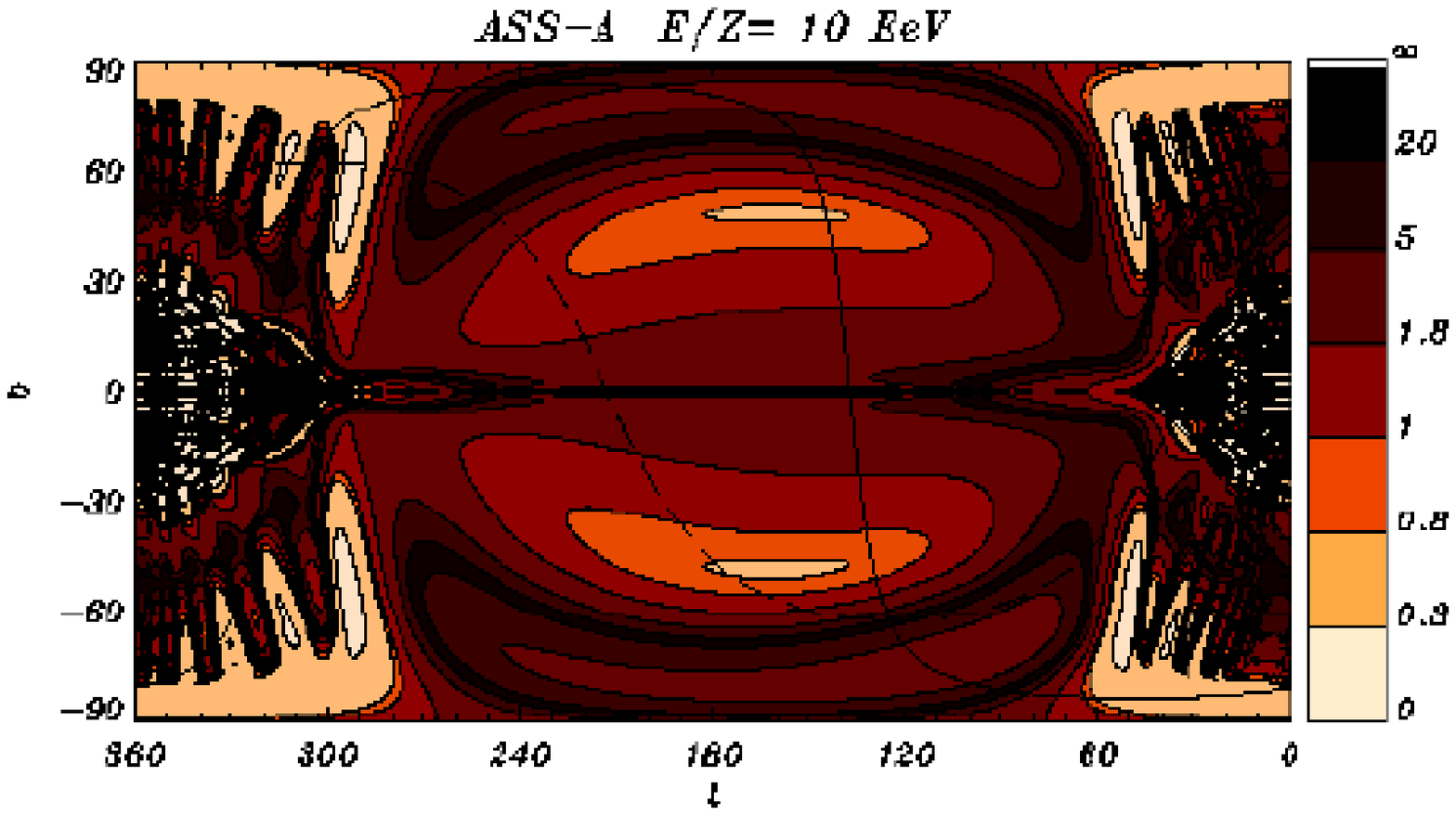,width=12cm}\caption{Same as
Figure 7, now in the ASS-A model and for $E/Z=10$~EeV.}}

Larger values of $\kappa$ than that estimated for the vertical 
direction may be expected for other
directions along which more intense field gradients are encountered 
or longer distances are traversed within significant field gradients.
Large (de)magnifications of UHECRs by the galactic magnetic field may 
thus occur when the ratio $E/Z$ is around and below a few tens 
of EeV, and even for larger values of this ratio in some special
directions. 

The precise determination of the amplification in the
galactic magnetic field requires the numerical integration
of eq. (\ref{disp}) coupled to eq. (\ref{mforce}).
We now turn to the numerical results of this calculation,
computed as described around eq. (\ref{disp}). 
Figure 7 shows contour plots of 
the magnification as a function of the observed arrival 
direction at Earth in galactic coordinates, for the 
BSS-S magnetic field model and for ratios $E/Z =$~30, 
10 and 5 EeV. Figure 8 is for the ASS-A model, at $E/Z=10$~EeV.
They were obtained through a $450\times 900$ grid which 
covered regularly all directions in the observer's plane.

The regions of very large amplification reveal
the existence of critical curves at which
the magnification formally diverges (something which
does never happen for realistic extended sources). 
It is 
instructive to compare these amplification contour plots
with the ``sky sheets'' shown in Figures 4 and 5. For instance,
the critical curves in the amplification maps at the observer's plane
correspond to the caustics in the source plane (the 
foldings in the sky sheets). Notice also that since the foldings
in the sky sheet separate regions of opposite image parity, 
the critical curves in the magnification maps must be closed
curves, as is apparent in Figures 7 and 8.

The value of the magnification at a given direction in
the maps of Figures 7 and 8 is the enhancement factor of the
flux that arrives to the Earth in that direction from a point source. 
Since the magnification is a function of energy
(for fixed $Z$) this effect changes the spectral slope
of the UHECR flux from point sources.

The ASS-A model  produces a magnification map symmetric
with respect to the galactic plane due to the odd nature of
the field with respect to the $z$-coordinate. An approximate
antisymmetry in the BSS-S magnification maps is apparent at
the largest value of the $E/Z$ ratio. At lower energies
this antisymmetry becomes less pronounced. It is  quite evident
that away from the galactic center, the regions in the northern
galactic hemisphere tend to be magnified while those in
the southern hemisphere tend to have significant demagnifications.
This can make sources with the same injected spectrum have a
harder observed spectrum in the South than in the North (i.e. steeper
in the North).

\subsection{Distortion of the spectrum of a point source}

We now illustrate through some particular examples
the dramatic changes upon the spectrum injected by a
point source that arise due to the combined effect of
flux (de)magnification and multiple image formation in
the galactic magnetic field. 

\bigskip

\FIGURE{\epsfig{file=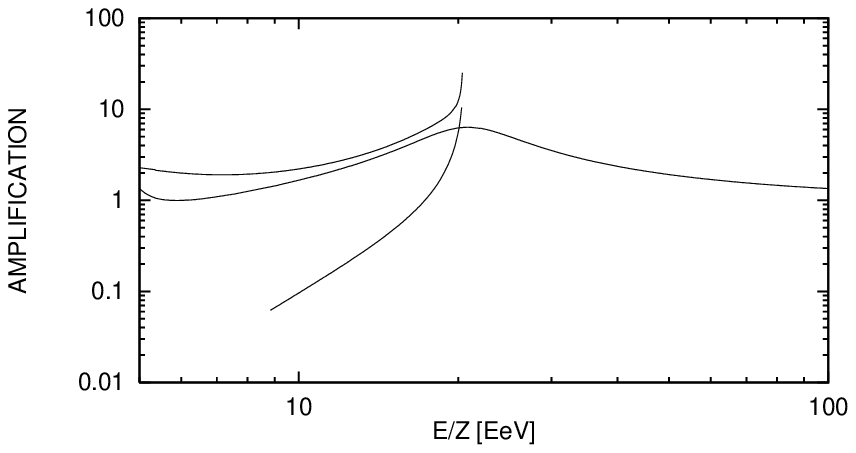,width=11 cm}
\epsfig{file=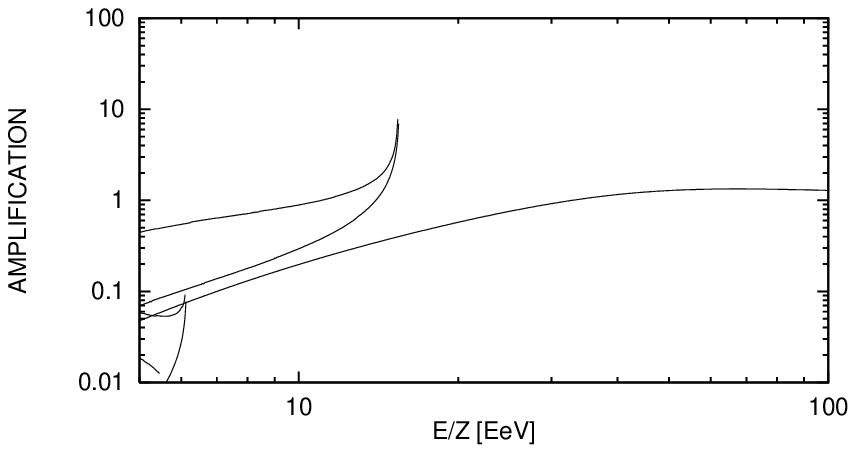,width=11 cm}\caption{Energy dependence
of the amplification factor
of the principal and secondary images of a source
at $(\ell,b)=(284^\circ,75^\circ)$ (M87, top panel) 
and at $(\ell,b)=(320^\circ,-30^\circ)$ (lower panel).
The magnetic field is taken in the BSS-S configuration.  
The spectra correspond to the same sources as in Figure 6.}}

Consider point sources at
the same two particular directions chosen in Section 3 
to illustrate the formation of
multiple images: $(\ell,b)=(284^\circ,75^\circ)$ 
(the Virgo cluster, visible from the northern terrestrial
hemisphere) and $(\ell,b)=(320^\circ,-30^\circ)$,
visible from the southern terrestrial hemisphere. 
Both lie next to the super-galactic plane. 
At the highest energies, $E/Z$ larger than 100 EeV, the CRs will
arrive to the Earth almost undeflected from the source direction and with
negligible magnetic lensing effects (amplification $\simeq 1$).
At lower energies the observed arrival directions change
and multiple images appear, as described in Figure 6
in Section \ref{images}. In Figure 9 we follow the amplification 
factor of each image as a function of the energy, for the BSS-S
galactic magnetic field model. The top panel corresponds to
the direction towards the Virgo cluster. The principal
image is largely magnified around $E/Z=20$ EeV, by a factor 
of order 5. At around that same energy  two new
highly amplified images form. One becomes rapidly
demagnified while the other remains magnified above
the source. In the example in the southern sky (bottom panel)
the principal image is steadily demagnified at energies 
below 30 EeV. Two highly magnified images form below
15~EeV but quickly become demagnified,  another
demagnified pair appears at energies below 6 EeV and
finally a new single and even more demagnified image
is found at still lower energies. 
Clearly the spectrum observed at the Earth from each of the 
images would be the
convolution of the injected spectrum with the plotted amplification.

\section{The Liouville theorem}

The Liouville theorem, i.e. the constancy of the phase space volume
along the particle trajectories, has important implications for the
effects we have been discussing here, making them essentially
undetectable for energies well below the ankle. This theorem was first
applied to CRs by Lemaitre and Vallarta \cite{le33} in their study of
the deflection of low energy (few tenths of GeV) 
CRs by the Earth dipole magnetic 
field. The conclusion \cite{le33,sw33} was that, due to the Liouville
theorem, an isotropic flux outside the solar system should remain
isotropic when observed at the Earth (except for the Earth shadowing
and the existence of blind regions corresponding to trajectories
leaving the Earth which
never manage to arrive to infinity \cite{le33}\footnote{Something 
similar can
happen to extra-galactic CRs, i.e. that when backtracking a CR out
from the Earth it remains trapped in the galactic magnetic field. No
extra-galactic sources will then be seen along these directions. This
will however happen only for $E/Z\lsim 1$~EeV, and would imply that
when projecting the ``sky sheet'' out from the Earth into the source
plane, one patch will remain trapped inside the Galaxy, leaving a hole
in the source sky.}). 
Since at low
energies the CR propagation throughout the Galaxy is diffusive, the CR
fluxes are essentially isotropic when  arriving to the solar system
and the theorem applies. In simple terms, when CR fluxes are
magnified, the angular spread of their velocities increases
 (a well known problem for
accelerator designers), so that they are seen as arriving from a
larger solid angle. When measuring the flux per unit solid angle, the
two effects compensate each other and an isotropic flux will
remain isotropic.
In gravitational lensing theory, this is also the reason why no
anisotropies can be generated in the CMB radiation by lensing effects
alone, which can only modify already existing anisotropies. 

The magnification effects we have discussed before would rather have
their gravitational analogue  in the spectacular observation of strong
lensing effects in quasars (multiple images) and also of microlensing
observations in the Galaxy and the LMC (huge magnifications, caustic
crossing due to binary lenses, etc.). Although with the present
limited statistics the CR sky
is not inconsistent with being isotropic, it is not expected to be so
at the highest energies, where CR should arise from a few very
violent sources (all with different intensities) and hence the Liouville
theorem will not hide the effects analyzed above.

We note that, thanks to the Liouville theorem, there is an alternative
way to compute the amplification for a given direction, besides the
intuitive one explained before. It consists in following three nearby
trajectories leaving the Earth with slightly different directions and
the same energy. If the initial velocities are  $\vec{v}_0$,
$\vec{v}_0+\Delta \vec{v}_i$ ($i=1,2$), one may define a solid angle
associated to those directions through d$\Omega=(\Delta
\vec{v}_1\times \Delta\vec{v}_2)\cdot\vec{v}/|\vec{v}|^3$. The
magnification can then be equivalently computed as d$\Omega_E/{\rm
d}\Omega_H$, i.e. as the ratio of solid angles at the Earth and at the
point where the particles leave the Galaxy ($r=20$~kpc). We have
verified that the two procedures agree in all directions to better
than a percent for $E/Z>5$~EeV, and this gives us further 
confidence in the
numerical accuracy achieved in the computation of the amplifications.

\section{Conclusions}

We have shown that the regular component of the galactic
magnetic field acts as a giant lens for UHECRs. It can lead to
multiple images, significant magnifications and deflections of the CR
trajectories. These effects are important for $E/Z\lsim 50$~EeV, and
hence can be relevant even for the highest energy events observed
($E\simeq 200$--300~EeV) if the CRs have a component which is not
light ($Z\geq$ a few).  Below $E/Z\simeq 1$~EeV, the trajectories
become so tangled that this description is no longer useful, and also
the galactic contribution to CR fluxes becomes dominant.

Actually, the same kind 
of magnetic lensing effect that we
discussed here  in detail for our Galaxy will
also affect the CRs at their exit from the
source galaxies. Hence,  even if the deflections at the source galaxy 
will not significantly
change the arrival direction to our galaxy (since the sources are far
away), they can distort the original spectrum produced in the
acceleration process.

We have emphasized the need to know the CR composition to do astronomy
even at the highest energies. Moreover, a better knowledge of the
galactic magnetic field is required. Besides establishing its general
features, such as the existence of reversals in the spiral arms, the
parity of the field across the galactic plane, the existence of an
halo component and its precise scale height, also a better
understanding of the magnetic field in the central parts of the Galaxy 
and whether a vertical component 
($B_z$ of a few tenths of $\mu$G, induced e.g. by a galactic wind) 
actually exists would be important. In this sense, the results
presented in this paper should be taken as indicative of the effects
that are expected in realistic models, although the detailed
predictions are model dependent.

Also the presence of a small scale random component could somehow
affect the predictions, making the magnetic lens a not perfectly
`polished' one and then blurring the images.
 It is conceivable that with the large statistics
expected with new detectors such as Auger \cite{auger}, the
identification  of point sources and
their study could also be helpful to understand the magnetic fields
themselves. 

The lensing effects we have discussed can have several
implications. For instance, the study of the CR spectrum in a region
around an observed source may show some of the features due to the
magnification which were illustrated in Figure~9. These could be
enhancements of the spectrum for energies leading to
 high magnification or
when  new images appear in
a caustic. Also, some regions of the southern sky appear  more
demagnified at low energies (for the reference BSS-S model), while
sources in the northern sky behave generally in the opposite way. This
can lead to harder spectra observed in some regions of the southern
sky and softer spectra in the north. This would be observable (with
some care) if not too many sources dominate the sky, since otherwise
sources from the north will be seen in the southern sky at low
energies, softening the average spectrum seen in the south.
It is also possible that the lensing effects could help in 
 the observation of
very faint sources, if they lie in regions of significant magnification.

\bigskip

\acknowledgments

Work partially supported by CONICET and Fundaci\'on Antorchas, Argentina.

\end{document}